# A Novel Modeling Approach for All-Dielectric Metasurfaces Using Deep Neural Networks

**Sensong An[1], Clayton Fowler[1], Bowen Zheng[1], Mikhail Y. Shalaginov[2], Hong Tang[1], Hang Li[1], Li Zhou[1], Jun Ding[3], Anuradha Murthy Agarwal[2], Clara Rivero-Baleine[4], Kathleen A. Richardson[5], Tian Gu[2], Juejun Hu[2], Hualiang Zhang[1,*]**

*[1]Department of Electrical & Computer Engineering, University of Massachusetts Lowell, Lowell, Massachusetts, USA*
*[2]Department of Materials Science & Engineering, Massachusetts Institute of Technology, Cambridge, Massachusetts, USA*
*[3]Shanghai Key Laboratory of Multidimensional Information Processing, East China Normal University, Shanghai, China*
*[4]Lockheed Martin Corporation, Orlando, Florida, USA*
*[5]CREOL, University of Central Florida, Orlando, Florida, USA*

*[hualiang_zhang@uml.edu](mailto:hualiang_zhang@uml.edu)

**Abstract**

Metasurfaces have become a promising means for manipulating optical wavefronts in flat and high-performance optical devices. Conventional metasurface device design relies on trial-and-error methods to obtain target electromagnetic (EM) response, an approach that demands significant efforts to investigate the enormous number of possible meta-atom structures. In this paper, a deep neural network approach is introduced that significantly improves on both speed and accuracy compared to techniques currently used to assemble metasurface-based devices. Our neural network approach overcomes three key challenges that have limited previous neural-network-based design schemes: input/output vector dimensional mismatch, accurate EM-wave phase prediction, as well as adaptation to 3-D dielectric structures, and can be generically applied to a wide variety of metasurface device designs across the entire electromagnetic spectrum. Using this new methodology, examples of neural networks capable of producing on-demand designs for meta-atoms, metasurface filters, and phase-change reconfigurable metasurfaces are demonstrated.

Metasurface devices or meta-devices, enabled by arrays of meta-atoms, provide a novel platform for realizing ultrathin and planar/conformal electromagnetic (EM) components and systems (e.g. meta-optics). By harnessing the electromagnetic multipoles excited within subwavelength meta-atoms (*1, 2*) that act as scattering particles, independent phase (and amplitude) control can be achieved at the element level by manipulating the geometry of the individual meta-atoms, allowing EM wavefronts to be tailored with high precision (*3-6*) for use in optical meta-devices. Recently, metasurfaces consisting of all-dielectric meta-atoms have drawn enormous attention, since they readily support magnetic multipole resonances and are less lossy at optical and infrared wavelengths compared to their metallic counterparts (*4, 7-11*). The multipole responses in a given meta-atom can be highly complicated even for simple shapes, and thus a meta-atom's impact on the phase and amplitude of an EM wave is difficult to predict. This problem is further complicated by the fact that meta-atoms are generally utilized in arrays and so accurate predictions must also account for the collective response of the array. As such, it can be time-consuming and laborious to find an appropriate set of meta-atoms for a particular design. Hence for these reasons reliable and efficient modeling tools are being heavily investigated. One approach is to develop analytical effective medium models, such as the Lewin model (*12*) and the GEM model (*13*). However, these models assume the long-wavelength approximation and become inaccurate when the wavelength is comparable to the meta-atom size. Another approach, which relies on iterative numerical full-wave simulations (e.g. finite-element method (FEM), finite-difference time-domain (FDTD) method and finite integration technique (FIT)), is widely adopted today. This approach provides accurate device response predictions but is severely time consuming. Moreover, the design process largely relies on empirical reasoning or trial-and-error (*8, 9*), which is inefficient and often ineffective, especially when the problem is highly nonlinear.

To overcome these obstacles, we consider a machine-learning-based design approach. Recently, machine learning (ML) has emerged as a powerful computational tool that has been broadly applied to many areas of science and engineering. It provides a promising solution to reducing time-consuming calculations and producing results with limited computational resources (*14-17*). Among the different machine learning techniques, deep neural network (DNN) based approaches have shown great promise for solving non-intuitive problems. DNNs usually contain multiple hidden-layers that provide sufficient hidden units, which can be used to represent complicated functions according to the universal approximation theorem (*18-20*). Therefore, it is possible to uncover hidden relations between variables, such as between nanophotonic structure geometries and their electromagnetic (EM) responses. Inspired by this idea, a data-driven method targeting rapid design by prediction of the EM responses of sub-wavelength structures has recently surfaced. Specifically, several DNNs that connect nanophotonic structures to their EM responses are constructed and then trained with a massive amount of pre-simulated data calculated by full-wave simulations based on FEM, FDTD or FIT. Recent progress shows that fully trained DNNs and inverse DNNs are able to predict the EM responses of select nanophotonic or metasurface devices on the scale of milliseconds (*21-23*). There are, however, three key standing challenges facing current DNN implementations: (1) dimensional mismatch between input and output data (*22*); (2) poor phase prediction accuracy; and (3) extension of DNN-based approaches from 1-D and 2-D structures to 3-D dielectric metasurfaces.

The first challenge arises because typical meta-atoms can be characterized using a limited number of variables (the inputs), but the output describing the response over a bandwidth of frequencies must be sampled at a sufficiently high rate to account for narrowband resonance features. Hence, a refined large-size tensor is required to accurately represent the transmission spectra, leading to a much larger number of outputs than inputs. To tackle this problem, one common approach is to shrink the size of the output tensor by sampling and interpolating (*21, 24*). However, this approach fails to provide accurate predictions of transmission spectra involving optical resonances, since the typical mean square error (MSE) loss function used in the regression task averages the error over the entire target frequency band and error contributions by resonance features are diluted and ignored. In another approach adopted in (*22*), an auxiliary network was created to predict the spectra around resonances at the expense of significant extra neural network construction and training efforts.

Although previous DNN-based approaches have yielded promising results in predicting amplitude responses of certain metasurface or nanophotonic devices (*17, 21-24*), accurate phase prediction has yet to be demonstrated. This difficulty is mainly caused by the 180° phase discontinuities that are introduced by the electromagnetic dipoles and/or quadrupoles (and possibly even higher order poles) inside the meta-structures (*25*). The only phase predicting DNNs presented in previous work (*21*) are limited by a relatively large average prediction error of 16°. Phase prediction failures severely restrict the accuracy of designing phase-based meta-devices such as meta-lenses or deflectors using DNNs or related approaches. Moreover, EM responses of all-dielectric 3-D metasurfaces/meta-atoms (which can support both magnetic and electric dipole resonances to obtain highly transmissive, full $2\pi$ phase coverage (*25*)) are difficult to predict, because the resonances excited within the 3-D structures all contribute to the scattering field with varying strengths (*10, 26*). As such, previous DNN implementations have been limited to either 2-D metallic metamaterials/metasurfaces or one-dimensional 1-D dielectric nanophotonic structures (*21-24*). 3-D dielectric meta-device design based on DNNs has been an open question.

In this work, a new approach to designing all-dielectric meta-devices employing DNNs is presented, which addresses all three key challenges discussed above. For the first time, predicting neural networks capable of simultaneously and accurately modeling amplitude and phase responses of all-dielectric meta-atoms over a wide spectrum have been demonstrated. Based on this highly accurate forward predicting neural network, several inverse design networks corresponding to different design objectives were constructed to illustrate the versatility of the method. These included a meta-filter design network, a meta-atom design network, and an index-reconfigurable meta-atom design network incorporating phase change materials. The design examples employing these networks substantiate that the proposed approach accomplished two important goals in the field of all-dielectric meta-device design: (1) reduction of time-consuming EM simulations for validating the performance of meta-device designs; and (2) finding non-intuitive device designs based on pre-determined EM response requirements, especially for multi-functional device designs. Importantly, the proposed approach also validates the feasibility of objective-driven 3-D optical device design, which can be easily generalized to many other electromagnetic problems, such as all-dielectric antenna and photonic integrated circuit designs.

## Results

**"Forward" predicting neural network.** To realize the DNN-based optical design functions and address the two goals described above, a unique deep network called the predicting neural network (PNN) is constructed. The PNN is a fast data-driven DNN-based tool that is able to predict the complete EM response (both phase and amplitude) of 3-D all-dielectric meta-structures with high accuracy. It plays a pivotal role in designing meta-devices with on-demand functions (as will be demonstrated later). The all-dielectric meta-device under consideration consists of a dielectric meta-atom (preferably with a high refractive index) and a dielectric substrate (preferably with a lower refractive index). Without loss of generality, cylinder-shaped dielectric meta-atoms are investigated as the first example due to their robust shape and low fabrication complexity (*10, 11*). During the modeling process, the meta-atoms are arranged in rectangular lattices, while their electrical permittivity, radius, height and the gaps between adjacent meta-atoms are considered as design variables. The spectra of interest are set in the infrared regime from 30 to 60 THz (5 μm to 10 μm in wavelength) for the purpose of demonstration.

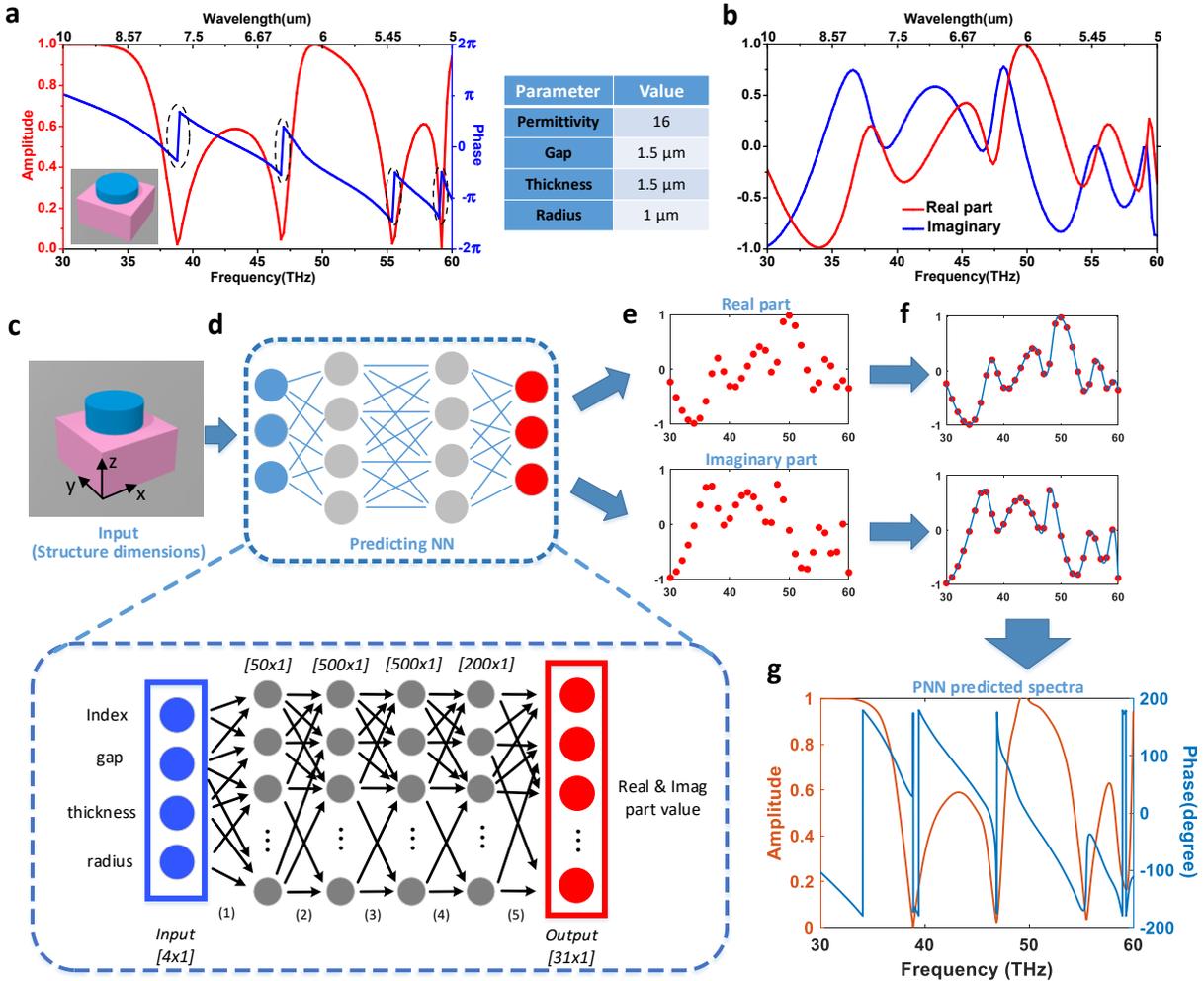

**Fig. 1. Methods and structure of the PNN. a** Numerically simulated amplitude (red) and phase (blue) responses of a sample all-dielectric meta-atom. The design parameters are listed in the accompanying table. The phase jumps (black dashed circles) represents the π phase discontinuities introduced by electromagnetic poles. **b** Real part (red) and imaginary part (blue) of the field transmission coefficient of the same meta-atom. **c** Input of the PNN, including the structural dimensions and material properties. **d** The main structure of the proposed PNN. Two individual fully-connected neural networks, each containing four hidden layers, are constructed to predict the real and imaginary parts of the complex transmission coefficient, respectively. Blue circles represent the input parameters, gray circles represent the hidden neurons, and red circles are the output values. **e** One sample of discrete real and imaginary part data given by the PNN. **f** Continuous data (blue curves) reconstructed by interpolating the discrete output samples (red dots). **g** The corresponding transmissive phase and amplitude.

Here we propose an implicit way to construct and train the networks to predict the amplitude and phase responses of meta-structures. For a typical meta-structure, like the one shown in Fig. 1a, the amplitude and phase response of its transmission coefficient will abruptly change (especially the phase) around resonant frequencies. In contrast, the real and imaginary parts remain smooth and non-singular (Fig. 1b). This observation prompted us to choose the prediction targets as the real and imaginary parts of the transmission coefficient, rather than the phase and magnitude during network constructing and training. This choice significantly improves the phase prediction accuracy by the network. In addition, the amplitude dips and phase discontinuities in Fig. 1a only happen when both the real and imaginary parts are close to zero (*27*). This enables us to down-sample the frequency points and reduce the output tensor dimension, which solves the mismatch problem. Considering that both phase and amplitude can be predicted through these two networks simultaneously, no further complexity was added.

The detailed architecture of the proposed PNN is shown in Figs. 1c-g. It deals with the regression problem between structure dimension parameters (refractive index, gap, radius and height) and the complex transmission coefficient over the 30-60 THz band. The whole spectrum is down-sampled into 31 frequency points (between 30 to 60 THz) with a frequency step of 1 THz, corresponding to 31 coefficients which are specified as the network output. Two independent neural networks (to be further discussed later) are constructed to predict the real and imaginary parts of the transmission coefficient, respectively. After the real and imaginary parts are derived, the amplitude and phase responses can subsequently be retrieved by applying the following equations:

$$Amplitude = \sqrt{Imag(S_{21})^2 + Real(S_{21})^2} \tag{1A}$$

$$Phase = \tan^{-1}\frac{Imag(S_{21})}{Real(S_{21})} \tag{1B}$$

Due to the high nonlinearity of the problem, we used a revised Neural Tensor Network (NTN) (*28*) in the predicting neural network instead of traditional fully connected layers (which take simple linear combinations of the previous tensor and pass them on to the next layer). More specifically, we replace the first standard linear neural network layer with a bilinear tensor layer that directly relates the two entity vectors across multiple dimensions. The output of this layer is given by:

$$Output = f\left(e^T W_1^{[1:k]} e + e^T W_2^{[1:k]}(e \odot e) + V\begin{bmatrix} e \\ (e \odot e) \end{bmatrix} + b\right) \tag{2}$$

where $f$ is the rectified linear unit activation function applied element-wise and $e$ is the vector of four design parameters (radius, height, etc.). $W^{[1:k]}$ is a $4 \times 4 \times k$ tensor and the bilinear tensor products $e^T W_1^{[1:k]} e$ and $e^T W_2^{[1:k]}(e \odot e)$ both result in a vector where each entry is computed by one slice of the tensor: $e^T W_1^{[i]} e$ and $e^T W_2^{[i]}(e \odot e)$, $i = 1, \dots, k$. $V$ and $b$ are parameters in the standard form of a neural network with dimensions of $k \times 8$ and $k \times 1$. Compared to standard neural networks where the entity vectors are simply concatenated, the main advantage of this NTN is that it can relate the two inputs multiplicatively instead of only implicitly through nonlinearity. For example, the permittivity ($index^2$) and cross sectional area component ($\pi r^2$) of a meta-atom can be given directly by $e^T W_1^{[1:k]} e$, while the volume component ($\pi r^2 h$) can be given by $e^T W_2^{[1:k]}(e \odot e)$. In contrast, it takes several concatenated layers for standard neural networks to generate these nonlinear components. Considering that these physical quantities are closely related to the meta-atom's EM responses, these additional vector interactions significantly accelerate the training process. In our design, $k$ is set to be 50, and this bilinear tensor layer is followed by three fully connected hidden layers containing 500, 500 and 200 neurons, respectively. The output layer contains 31 units, corresponding to the 31 frequency points that were chosen to sample the 30-60 THz spectrum under consideration. The supervised training process is performed by minimizing the loss function, which measures the squared differences between the spectra prediction generated from the network and simulation results given by full-wave EM simulations.

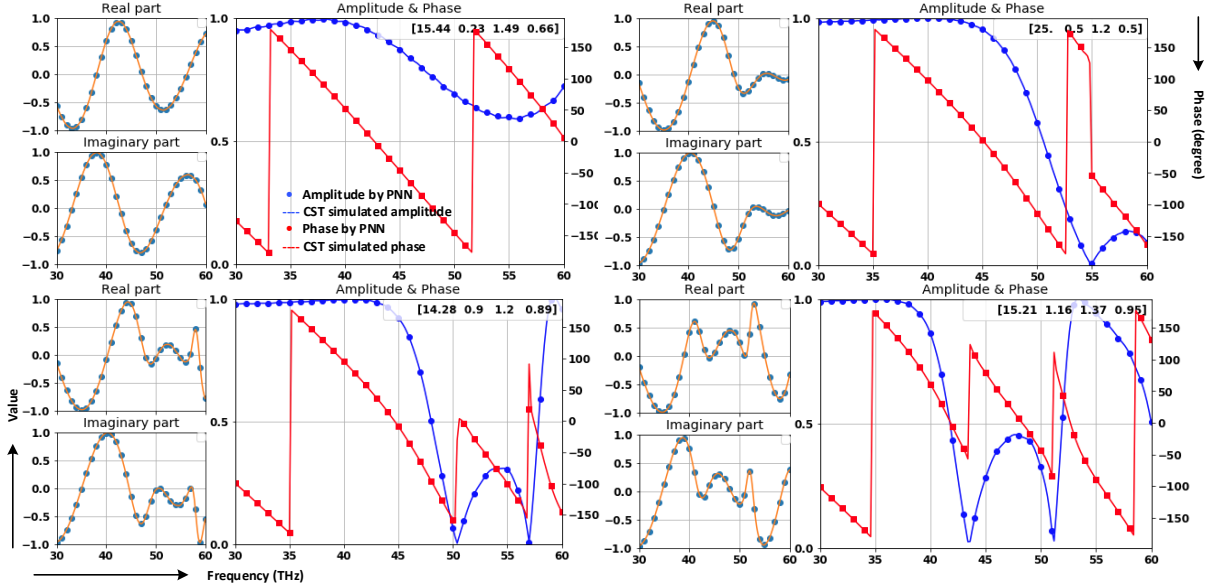

**Fig. 2. Examples demonstrating the PNN performance.** Small subplots shown on the left are the real and imaginary parts of each meta-atom's transmission coefficient. The red curves shown in the large subplots represent the phase profiles, while the blue curves refer to the amplitude responses. Dots represent data generated by the PNN, while solid curves are data obtained from numerical simulations. Design parameters of each meta-atom are given in the insets in the following order: permittivity, gap, thickness and radius (in microns). All four meta-atoms presented are randomly selected from the test data.

Over 50,000 groups of randomly generated $1 \times 4$ input vectors were fed into the bilinear tensor layer. Among them, 70% are assigned to the training set, while the remaining 30% are used as the test set. We then calculated the predicted real and imaginary parts and compared the results to full-wave simulations to extract the prediction error. After the training is completed, the overall test mean square error is 0.00035 for the real part, 0.00023 for the imaginary part, with a corresponding fractional error of 0.5% for both amplitude and phase responses (see methods for error definitions). Several prediction samples randomly selected from the test data sets are showcased in Fig. 2 (insets show the corresponding design parameters). These examples show excellent consistency between numerical simulations and PNN predicted results across the full spectrum under consideration. Given the vital role the PNN plays in our deep-learning approach, such agreement is critical to objective-driven meta-device design as will be discussed in the following sections.

**On-demand meta-atom design.** To demonstrate the efficacy of our deep learning design approach, we will first apply it to designing meta-atoms, the fundamental building blocks of meta-devices. In general, functionalities of most meta-optical devices, such as beam deflectors and lenses, are achieved by tailoring the wave front using a group of meta-atoms that cover the full $2\pi$ phase range (*29*). Therefore, a meta-atom design method capable of rapidly and precisely identifying design parameter combinations with large phase coverage is highly advantageous compared to time-consuming design iterations based on full-wave simulations. However, due to the diverse meta-atom design goals and restrictions on input parameters, it is unrealistic to build one single design network that applies to all situations. For example, most meta-atom designs stipulate a uniform refractive index, lattice size, and height, but allow varying radii (*10, 11, 25*). In contrast, fixed-gap meta-atom designs specify the size of gaps (*30, 31*). Even refractive index can become a design variable for meta-atoms based on phase-change materials (*32-34*). To meet the different application-specific needs, we adopt a closed loop design network instead of constructing a single cascaded inverse design neural network. Fig. 3a shows the architecture of the constructed meta-atom design network. The output design parameters given by the meta-atom model generator are fed to the PNN, and the predicted transmission and phase spectra of the current design are then sent back to the model generator, where a new design is given to further reduce the difference between the current result and the final goal. The model generator can be readily modified to meet different design requirements. As a demonstration, we constructed a meta-atom design network that produces meta-atom designs with large phase coverage. Here the operating frequency and permittivity are fixed and specified as inputs, while the network explores the parameter space to find the period and thickness combination that provides the largest phase coverage (ideally $2\pi$) with maximum transmittance.

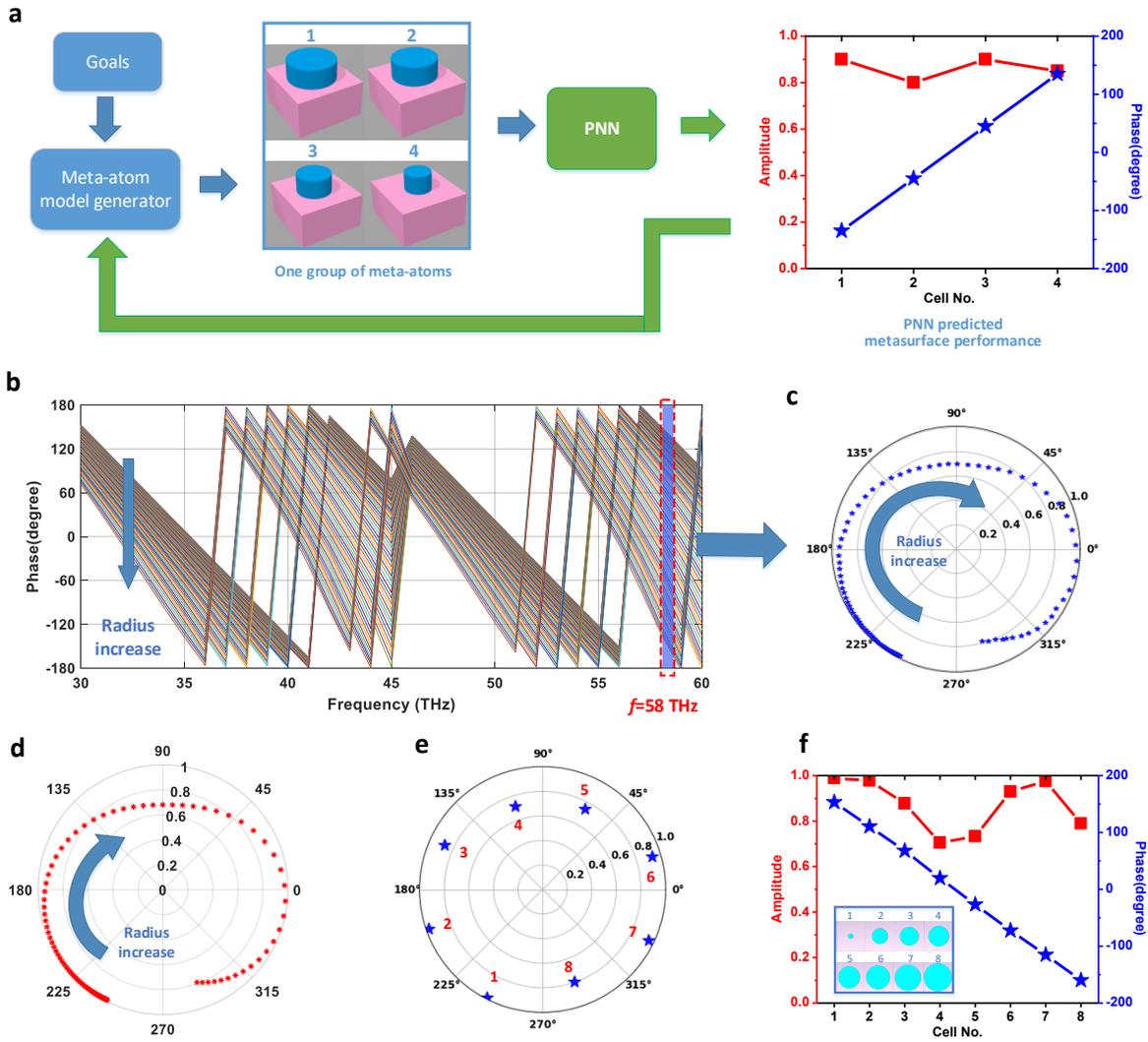

**Fig. 3. Structure of the proposed meta-atom design network and a design example. a** Flowchart of the closed-loop meta-atom design network. Designs produced from the model generator are evaluated by the cascaded PNN, and the predicted EM responses shown in the rectangular plot are subsequently sent back to the design generator. New designs are then given to minimize the differences between the current results and design goals. **b** PNN-predicted phase profiles of proposed designs over the spectrum. All 93 designs have 1.23 μm lattice size and 1.5 μm thickness, while their radii vary from 0.1 μm to 0.56 μm in 5 nm increments. **c** PNN-predicted phase and amplitude profiles at the target frequency (58 THz). **d** Numerically-simulated results produced for verification. **e** Eight cells picked from the simulated data with different radii (100, 315, 370, 410, 445, 475, 505 and 560 nm) to form a class of 3-bit meta-atoms (with a phase step of 45 degrees). **f** Performance and top-views (inset) of these eight cells.

In this specific example, the permittivity and operating frequency are fixed to be 24 and 58 THz respectively, while the upper bound of the meta-atom thickness is set to be 1.5 μm due to a presumed fabrication constraint. A minimum amplitude transmission threshold of 0.7 is also included in the generator to ensure high optical efficiency. With the help of design algorithms (e.g. genetic algorithm), the model generator searches for the optimized lattice size and thickness combination that can provide the largest phase coverage, which is 1.23 μm and 1.5 μm, respectively. The corresponding phase coverage is 330° when the meta-atom's radius is changed from 0.1 μm to 0.56 μm. A total of 93 samples, with a radius step of 5 nm, are chosen to sketch the phase coverage range in Fig. 3b. Phase and amplitude profiles at the working frequency (58 THz) are sliced and shown in Fig. 3c. Full-wave simulation results are also included in Fig. 3d for comparison. The excellent agreement between Figs. 3c and 3d once again shows the accuracy of the PNN. The 330-degree phase coverage above the preset amplitude threshold enables the design of 3-bit meta-atoms (a set of 8 meta-atoms), with a 45-degree phase difference between adjacent cells as depicted in Figs. 3e and 3f. It only took 22 seconds (see Table S2) to find the optimal design parameters using the proposed approach, while these designs' numerical verification alone (shown in Fig. 3d) took more than 30 minutes with the same hardware and environment settings. Arriving at this design via conventional methods would have taken hours or days depending on the skill and luck of the designer.

To further demonstrate that the developed meta-atom design network constitutes a universal design methodology, the approach is used to explore meta-atoms with a tunable refractive index for use in reconfigurable meta-device designs. The design network presented in Fig. 4a has been slightly modified by removing the permittivity from the input vector and adding meta-atom radius to the output vector. The general goal remains the same: find the optimal design with maximum phase coverage. The results are presented in Fig. 4. In this example, the index tuning range was set from 3.5 to 4.5 with reference to the widely-applied phase change material GST (GeSbTe) (*34*) while similarly assuming an operating frequency of 58 THz. According to PNN predictions, the generated optimal design (1.04 μm gap, 0.79 μm radius and 0.91 μm thickness) is able to achieve more than 320 degrees of phase coverage while the meta-atom switches progressively from one state to the other. 101 samples with a refractive index increment of 0.05 are chosen to illustrate the phase coverage range in Fig. 4a. Corresponding numerical simulation results are also included in Fig. 4b. Four meta-atoms with the same shape but different indices (Fig. 4c) are selected from these 101 individual designs to form a set of a 2-bit meta-atom design. Figs. 4d and 4e sketch the field distributions inside the meta-atoms. In state 1, which corresponds to a material index of 3.57, the meta-atom supports an electric dipole resonance as shown in Fig. 4d. When switched to state 4 with a material index of 4.15, a strong magnetic dipole moment emerges as evidenced by the field profile in Fig. 4e. The alternating electric and magnetic dipole resonances with changing index of refraction is essential to realizing full $2\pi$ phase coverage, and our meta-atom design network precisely captures this critical feature to identify the design parameters within a short timeframe (less than 5 minutes, details can be found in Table S2). In contrast, conventional methods necessarily demand laborious parameter sweeps and elaborate field distribution analysis to find a design fulfilling the same function (*25, 35*).

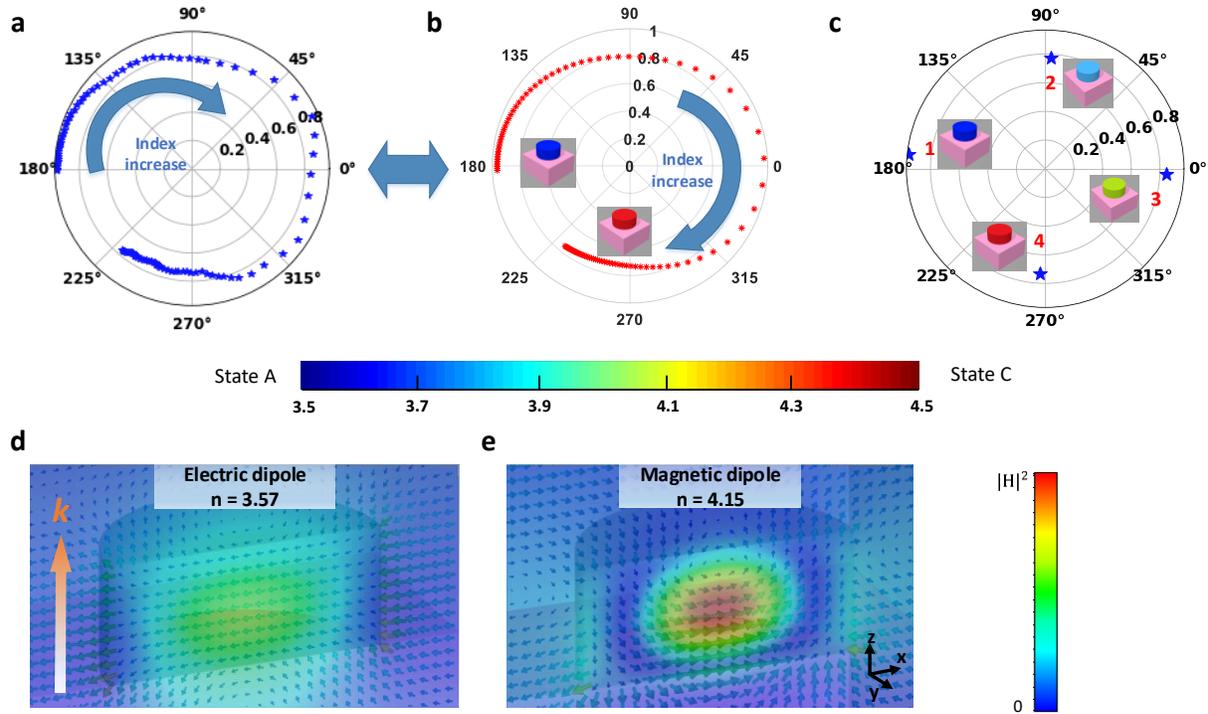

**Fig. 4. A 2-bit reconfigurable meta-atom design example. a** PNN-predicted phase and amplitude profiles of the optimized reconfigurable meta-atom design. All 101 meta-atoms have the same dimensions: 1.04 μm gap, 0.79 μm radius and 0.91 μm thickness, while their refractive indices vary from 3.5 to 4.5 with a 0.01 spacing, to represent intermediate states of a material phase change. **b** Numerically-simulated phase and amplitude results of designs generated by the meta-atom design network. Insets schematically depict meta-atoms in different material phases. A color bar correlating the colors and index values is shown at the bottom. **c** Four meta-atoms (insets) selected from (a), with refractive indices of 3.57, 3.89, 3.99 and 4.15, to form a class of 2-bit meta-atom design with 90-degree phase increments. **d, e** Electric (arrows) and magnetic (color coded) field distributions inside the meta-atom under (d) state #1 and (e) state #4. Arrow showing in the left indicates the incident light direction.

**"Inverse" neural network enabled on-demand meta-filter design.** Meta-filters or frequency-selective surfaces (FSS) represent another widely-used class of meta-devices. The design objective of meta-filters is a pre-assigned target transmission spectrum which can be parameterized as a vector, suggesting that the model generator can also be constructed with a DNN – the "inverse design" deep neural network. Once the data set is created and the inverse DNN is trained, the design progress is non-recurring and the model generators are inquired only once per design target. Therefore, this approach is extremely time-efficient.

The meta-filter design network is illustrated in Figs. 5a-e. In principle, a fully functional meta-filter design network should be able to generate meta-atoms with performance resembling the user-defined filter spectral responses. To achieve this goal, we connected the fully-trained PNN to a meta-filter generator to form a cascaded network and avoid the non-convergence problem resulting from non-unique solutions (*21, 22*). The meta-filter generator is also a DNN (aka the "inverse" network) that consists of four consecutive fully-connected hidden layers containing 500, 500, 500 and 50 neurons, respectively. As shown in Fig. 5a, the meta-filter generator employs the target spectrum as the input. In the output layer, a 4 by 1 output vector is generated, which contains the parameters of the newly generated design. The design parameters are then designated as input for the consecutive PNN, where the design's electromagnetic response is evaluated. Finally, the transmission spectrum of the current design is compared to the target spectrum and the Euclidean distance between them is calculated. During training, the weights and biases in the hidden layers of the meta-filter generator are optimized to minimize this distance, while the values of hidden neurons in the previously-trained PNN remain unchanged. As a result, the model generator becomes "smarter" as training proceeds, eventually forming a cascaded DNN able to generate on-demand meta-filter designs on a one-time calculation basis.

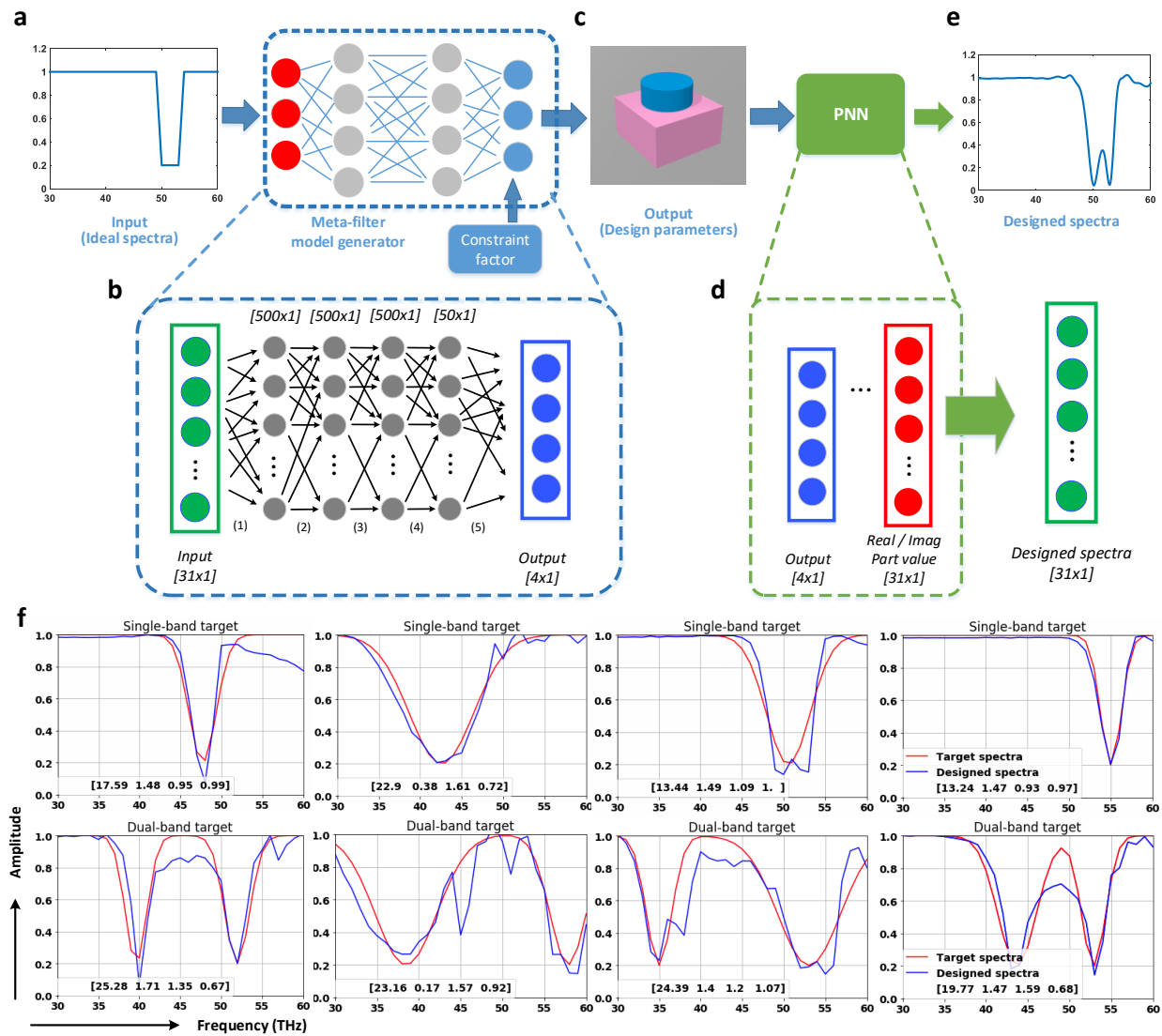

**Fig. 5. Architecture of the meta-filter design network and design examples. a** Target spectra designated as input. **b** The model generator of the meta-filter design network, which is constructed using DNNs. Cells in green represent the inputs, whereas gray and blue cells represent the hidden neurons and outputs, respectively. **c** Output of the model generator, which is a combination of design parameters such as permittivity and meta-atom dimensions. **d** These parameters are then fed into the PNN to yield the complex transmission coefficient. **e** The refined amplitude response of the generated design as derived from the complex transmission coefficient. **f** Several design examples from the proposed meta-filter design network on single-band design targets (first row) and dual-band design targets (second row). Red curves are target filter spectral responses, and the blue curves are the PNN-predicted filter spectral responses based on the designs given by the design network. All design parameters including permittivity, gap (μm), thickness (μm) and radius (μm) are given as insets.

Two data sets were used to train the meta-filter design network, including randomly generated Gaussian-shaped single-band and dual-band targets. Specifically, these targets were created using the following equations:

$$S_{single}(f) = 1 - 0.8\exp\left[-\frac{(f-f_0)^2}{2\sigma^2}\right] \quad (3)$$

$$S_{dual}(f) = 1 - 0.8\exp\left[-\frac{(f-f_0)^2}{2\sigma^2}\right] - 0.8\exp\left[-\frac{(f-f_0')^2}{2\sigma'^2}\right] \quad (4)$$

where $f_0, f_0'$ are the center frequencies of stopbands and $\sigma, \sigma'$ dictate the bandwidth. The final training sets include 20,000 groups of randomly generated single-band targets and 20,000 groups of dual-band targets, among which 80% are used for training and the remaining 20% are left for accuracy tests. The training was executed in an unsupervised way, since the input transmission spectra are not labeled with corresponding structure dimensions. After 50,000 epochs of training for each group of data, the error (see Methods for definitions) eventually stabilized at 15.0% and 24.7% for single-band targets and dual-band targets, respectively. Since the randomly-generated target spectra may be physically unrealistic, these stabilized error values indicate that the training had completed. The relatively large error of single-band targets and even larger error for dual-band targets do not present a limitation to our DNN approach; rather they manifest the inherent inability of achieving increasingly complex filter functions with the simple cylinder-shaped all-dielectric meta-atom geometry.

**PNN vs. interpolation.** The PNN is a far more sophisticated, precise, and powerful tool compared to interpolation algorithms for two reasons. First, a well-trained PNN offers superior performance in making predictions based on the same prior information. To demonstrate this, we find a unique method to compare the PNN's performance with a built-in interpolation function from the numerical computing tool MATLAB. As shown in Fig. 6a, we first find the 27 groups of data (in blue circles) that are evenly distributed in the parameter range: gap $\in [1.2, 1.4]$, thickness $\in [1, 1.2]$ and radius $\in [0.6, 0.8]$ (all in μm), with a spacing of 0.1 μm for each parameter. Then we randomly chose two test parameter combinations within this parameter space (indicated by the green and red triangles) and predict their transmission coefficients with the PNN and the interpolation tool, respectively. Since these 27 groups are the only training datasets existing in this specific parameter space, the PNN and the interpolation tool have "equal knowledge" about data within this range. Using a spacing of 0.01 μm for each parameter, we created $21 \times 21 \times 21$ (a total of 9,261) query points (blue dots in Fig. 6b) and performed two types of interpolations (linear and cubic) for each query point (including the two test samples). According to the results shown in Figs. 6c and 6d, spectra obtained by the interpolation tools (blue squares and green triangles) are much less accurate compared to the PNN predictions, in particularly at the short wavelength (high frequency) end of the spectrum. In contrast, the PNN-generated spectrum (red circles) maintains high accuracy across the entire 30-60 THz band.

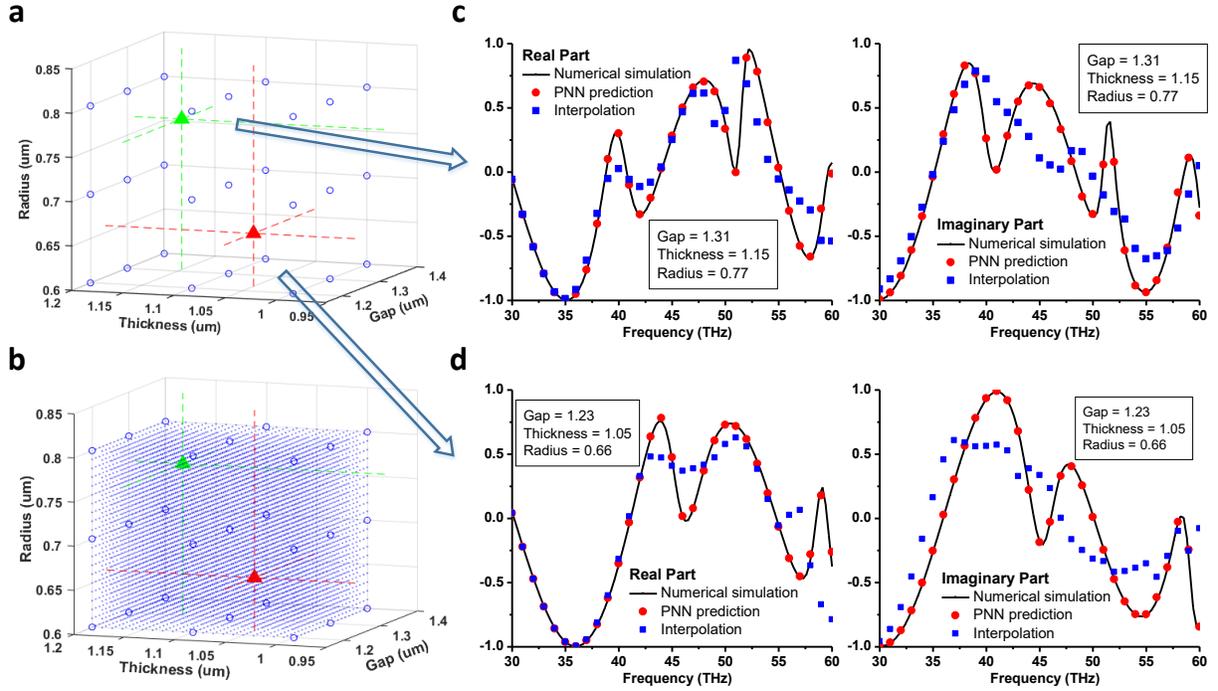

**Fig. 6. Comparison between PNN and data interpolation. a** The parameter scope chosen to perform the comparison. Blue circles indicate the 3 by 3 by 3 data points for interpolation. Green and Red triangles indicate the locations of the two test samples in this parameter scope. All data samples have the same dielectric constant of 24. **b** The meshed parameter scope with 21 by 21 by 21 query points. Both test parameter combinations are covered in these query points. **c** Prediction results with the PNN (red circles), linear interpolation (blue dots), cubic interpolation (green stars) and numerical simulation results (black lines) for the first test data point (green triangle in a and b) **d** PNN, interpolation and simulation results for the second (red triangle in a and b) test data point. Dimensions of these two test samples are given as insets.

The second reason is that the proposed PNN can also be used for extrapolation to make predictions outside the parameter space of the training data set. Ordinary extrapolation requires assumptions by a designer about the physical behavior outside of the data set (linear, polynomial, etc.), which are unlikely to hold over a significant data range. Moreover, the best choice of model may not be intuitively clear, particularly for multivariable problems. The PNN is better informed than a traditional designer when it comes to making accurate extrapolation predictions, because it can draw much more information from the training data to unravel intrinsic physical behavior of the system. We explored the PNN's "out-of-range" prediction capacity by feeding it with meta-atom structures with one or more parameters residing outside of the preset training data range. As shown in Fig. 7, the PNN retains excellent prediction accuracy when the inputs are not too far beyond the training data set boundaries. This interesting discovery indicates the DNN-based method's potential for uncovering the hidden physical mechanisms behind the large amount of input data. Nonetheless, since the PNN's performance deteriorates as the inputs move far away from the preset data range, it is important that certain boundaries of the collected data, such as fabrication limits, system requirements and design interests should be carefully determined before the network is constructed.

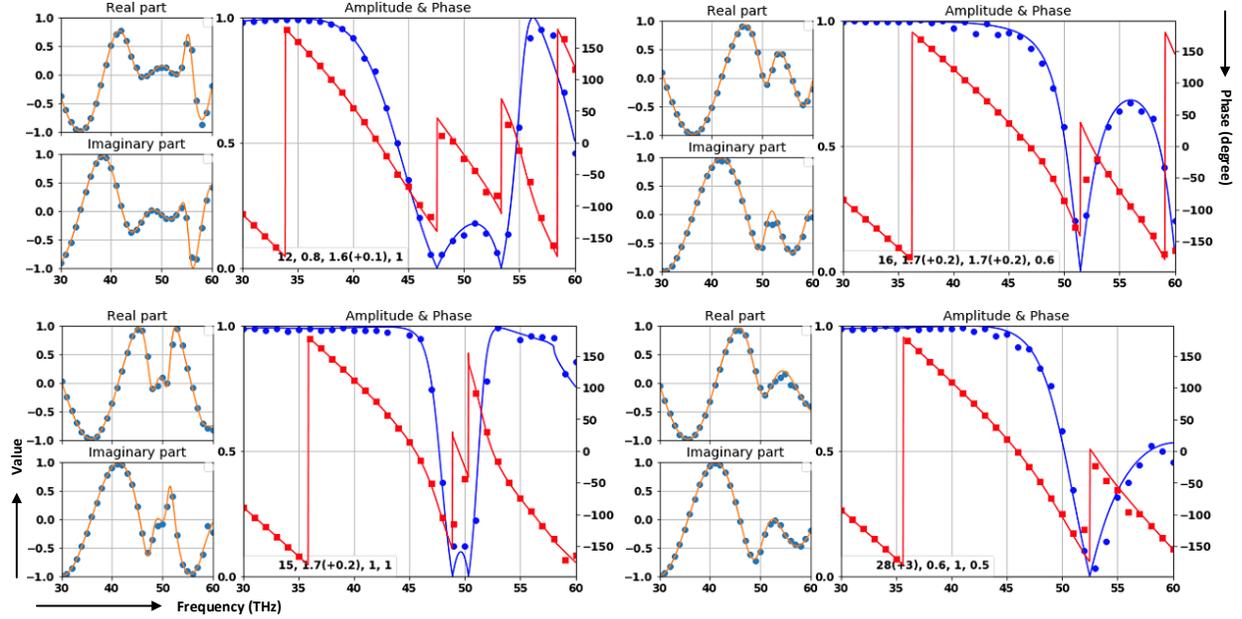

**Fig. 7. Examples of the proposed PNN with test data outside of the learning range.** Smaller subplots shown on the left are the real and imaginary parts of each meta-atom's transmission coefficient. The red curves shown in the larger subplots represent the phase profiles, while the blue curves refer to the amplitude responses. All dots represent data generated by the PNN, while solid curves are data obtained from the numerical simulation tool. Design parameters including the dielectric constant, gap (μm), thickness (μm) and radius (μm) of each meta-atom are given in the insets. All four meta-atoms presented have one or more parameters residing outside of the preset training data range. Values in the parentheses represent the distance from the preset data boundary.

**PNN for H-shaped meta-atoms.** Finally, to demonstrate that the method is universally applicable to meta-atoms of different geometries and not limited to the design of cylinder-shaped dielectric nanostructures, we trained another PNN for H-shaped meta-atoms (*4*). As shown in Fig. 8a, H-shaped structures can be uniquely determined by six parameters, which are combined into $1 \times 6$ vectors and assigned as the PNN's input. Dimensions of the bilinear tensor layer in the PNN are slightly modified to adapt to this change. The structure of other hidden layers and the output vectors are the same as those of the PNN for cylinder-shaped meta-atoms. After being trained with 14,800 groups of H-shaped meta-atom datasets, the PNN is able to achieve 99.4% accuracy for amplitude prediction and 99.5% for phase prediction. Three instances were presented in Figs. 8b-d to demonstrate the PNN performance, and more examples are included in Fig. 8.

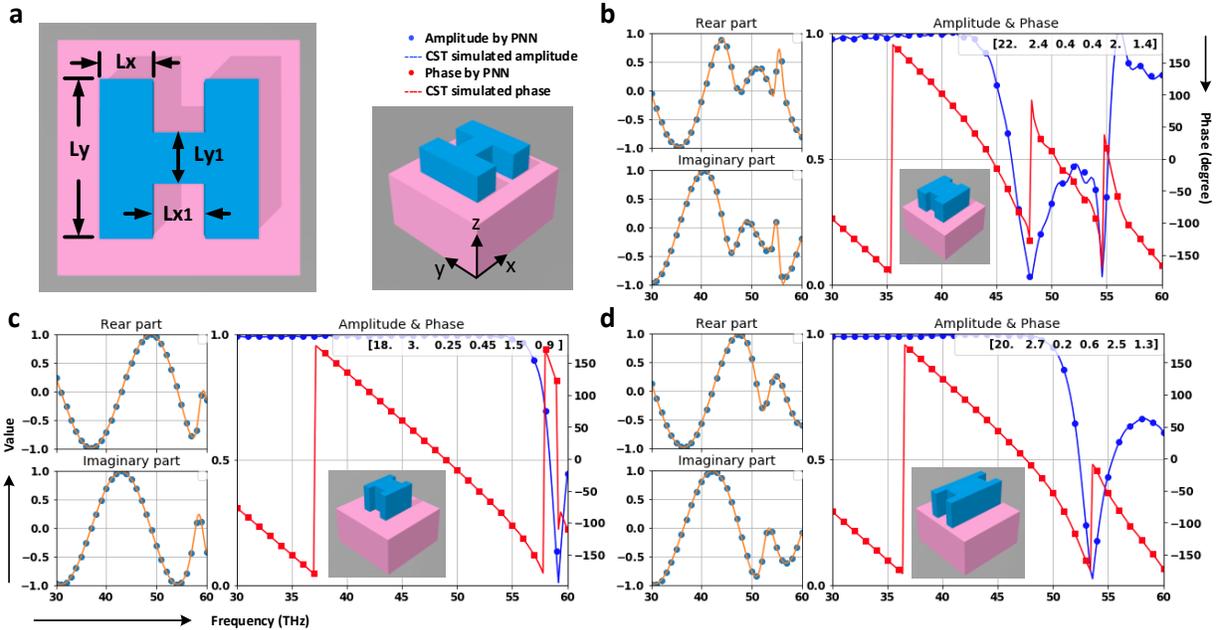

**Fig. 8. PNN for H-shaped dielectric meta-atoms. a** Top-view figure showing the dimensions that parameterize the H-shaped meta-atom structure. **b-d** Examples of EM responses of H-shaped meta-atoms predicted by the PNN (dots) and full-wave numerical simulations (lines). Meta-atom design parameters are given as insets in the following order: permittivity, lattice size, $L_x$, $L_{x1}$, $L_y$ and $L_{y1}$ (in microns).

## Discussion

The DNN-based design method features several noteworthy advantages. First, compared to traditional meta-device design approaches, a trained neural network avoids the large volume of full-wave simulations and realizes on-demand designs in a few seconds. By generating an optimal design that provides the closest match to the preset performance target, the network is able to quickly determine whether or not the target is physically viable based on the given meta-atom design constraints (e.g. sizes and geometries) and prescribe the performance limitations. This is accomplished without involving numerous computationally intensive, iterative full-wave simulations necessary to cover the enormous parameter space (e.g. permittivity, shapes, and geometric dimensions). Second, the DNN-based method can readily handle non-intuitive, multifunctional meta-device design. For example, it is intuitive to conceptualize a single-band meta-filter design, because larger meta-atom volume is usually correlated with lower resonant frequencies. The target stopband frequency can therefore be straightforwardly met by adjusting the radius and/or height of the meta-atom. However, for multifunctional meta-atom designs, changing one design parameter (radius, gaps or thickness) inevitably affects other functions (e.g. phase or amplitude response at another frequency or polarization direction). Escalating complexity resulting from such coupling makes iterative design purely guided by intuition nearly impossible. DNN-based methods are uniquely poised for solving such complex, multi-objective design challenges. Third, the PNN's remarkable accuracy is retained even with out-of-range parameters, which suggests that its network architecture is able to capture the intrinsic physical traits of the light-matter interaction process in the metasurface structures. With this unique property and unparalleled speed, the solutions provided by the well-trained design networks can potentially be used to elucidate the underlying physical mechanisms behind nanophotonic structures such as coupling and resonances.

Compared to previously proposed DNN-based optical device design methods (*17, 21-24*), the PNN realized in this paper resolves difficulties associated with predicting resonant frequencies and achieves wideband phase prediction for the first time. In addition, we successfully expanded DNN-based methods from 1-D and 2-D metamaterials to 3-D dielectric metamaterials. Importantly, our PNN approach is geometry-agnostic, evidenced by its successful application to H-shaped meta-atom design. Since the in-plane pattern of any lithographically defined meta-atom can always be parameterized with a 2-D matrix, our approach can be extended to metasurface designs based on topologically optimized, non-intuitive meta-atom geometries by using the 2-D matrix as the PNN input.

Like all other DNN-based approaches, performance of the PNN (and on-demand design networks) is limited by the quality and quantity of training data. The PNN is a data-driven simulator, which becomes increasingly accurate in finding the implicit yet inherent connections between meta-atom structures and their performances as more training data are included. The data collection process, which takes significantly longer time than network construction and training, is a bottleneck for this methodology. Nevertheless, this shortcoming can be overcome with data sharing: large amounts of simulation data have already been and are continuing to be generated during the traditional trial-and-error design process. These data, including what traditionally were considered as "failed" designs, constitute an existing asset that can readily be utilized to build training datasets. Moreover, once the training is completed, the on-demand design network is able to provide the optimal designs with minimal amount of time and is generically applicable to different design targets. This result is readily understandable from the standpoint that unlike numerical interpolation, DNN builds on the intrinsic (albeit non-intuitive and hidden) relation between metasurface structure and its electromagnetic response. The superior data-driven prediction efficacy positions DNN as the universal metasurface design optimization method of choice.

In summary, we have introduced a novel DNN-based data-driven approach for accurate prediction of all-dielectric meta-devices' responses, as well as inverse design of the meta-devices based on pre-defined performance targets. For the first time, amplitude and phase responses of all-dielectric meta-atoms are simultaneously derived in millisecond-timescale. We further show that objective-driven meta-device design models can be constructed based on this predicting neural network. Although this paper mainly discusses the all-dielectric meta-devices designs operating at the infrared spectrum, the deep learning method developed herein for objective-driven design is not limited to this context. In general, the proposed framework can be adapted to the design of other complex electromagnetic media such as multi-functional meta-atoms, integrated photonic devices, and optical antennas.

## Methods

**Data collection.** Figs. 1 and 8 illustrates the general schematic of the all-dielectric structures under consideration. They all consist of a dielectric meta-atom (upper layer) and a dielectric substrate (bottom layer). During the modeling process, the meta-atoms are arranged in rectangular lattices. Random parameter combinations including the gap, thickness, radius and permittivity of meta-atoms were generated in the multi-paradigm numerical computing tool MATLAB, and then transferred to commercial software package CST Microwave Studio for full-wave simulations. The parameters are created with (all lengths in microns): gap $\in [0.1, 1.5]$, thickness $\in [0.5, 1.5]$, radius $\in [0.1, 1.2]$, index $\in [3.5, 5]$, since these parameter ranges include ample samples of phase and amplitude responses. Real part and imaginary part data of the transmission coefficient over the operating spectrum are calculated using CST time domain solver, with the unit cell boundary condition applied for all meta-atoms in both $x$ and $y$ directions. For these meta-atoms, an *x*-polarized plane wave was illuminated from the substrate side. Open boundaries are implemented in both the negative and positive *z* directions (the axes are defined in Figs. 1 and 8).

**Network construction.** 35,000 groups of cylinder-shaped meta-atom models and their corresponding complex transmission coefficients are collected and used for the training of PNNs. The design generator in the meta-atom design network employs an evolutionary computation framework, DEAP, to generate new design parameters. The evaluation function of the framework is set to be the phase coverage of the current design calculated by the trained PNN. All DNNs models are constructed under the open-source machine learning framework of TensorFlow.

**Loss function & error.** The loss functions we use for PNNs are L2 loss functions, which stand for least square errors, also known as LS. More specifically:

$$L_{PNN} = \frac{1}{N} \sum_{i=1,2...N} (S_{prediction} - S_{simulation})^2 \qquad (5)$$

which measures the squared differences between the spectra prediction generated from the network ($S_{prediction}$) and the simulation results given by full-wave electromagnetic simulations ($S_{simulation}$).

For the meta-filter design network, practical material and fabrication limitations will likely constrain the choices of refraction index, thickness, gap, and radius of the meta-atoms. Therefore, a constraint factor is added to the loss function of the meta-filter design network. The revised loss function is defined as:

$$L_{Filter} = \frac{1}{N} \sum_{i=1,2...N} (S_{prediction} - S_{target})^2 + (P - P_{clipped}(max, min))^2 \qquad (6)$$

where $P$ is the output vector containing design parameters and $P_{clipped}$ is a vector with all values in $P$ clipped to a preset maximum and minimum value. The constraint factor $(P - P_{clipped})^2$ measures the distance from output to the desired parameter value range, when the vector $P$ falls out of the preset value range, the increasing loss function value $L_{Filter}$ will force the trainer to re-assign a $P$ value within the desired range $(max, min)$.

The errors we use to evaluate the training results are fractional differences, which are defined as:

$$E_{PNN} = \frac{1}{N} \sum_{i=1,2...N} \left( \frac{S_{prediction} - S_{simulation}}{S_{simulation}} \right) \qquad (7)$$

$$E_{Filter} = \frac{1}{N} \sum_{i=1,2...N} \left( \frac{S_{prediction} - S_{target}}{S_{target}} \right) \qquad (8)$$

# Reference

1. H. C. Hulst, H. C. van de Hulst, *Light scattering by small particles*. (Courier Corporation, 1981).
2. J. D. Jackson, *Classical electrodynamics*. (John Wiley & Sons, 2007).
3. J. Ding, S. An, B. Zheng, H. Zhang, Multiwavelength Metasurfaces Based on Single-Layer Dual-Wavelength Meta-Atoms: Toward Complete Phase and Amplitude Modulations at Two Wavelengths. *Adv. Opt. Mater.* **5**, 1700079 (2017).
4. L. Zhang, J. Ding, H. Zheng, S. An, H. Lin, B. Zheng, Q. Du, G. Yin, J. Michon, Y. Zhang, Ultra-thin high-efficiency mid-infrared transmissive Huygens meta-optics. *Nat. Commun.* **9**, 1481 (2018).
5. S. An, J. Ding, B. Zheng, Y. Lin, W. Zhang, H. Zhang, in *CLEO: Science and Innovations*. (Optical Society of America, 2017), pp. JW2A. 104.
6. X. Wang, J. Ding, B. Zheng, S. An, G. Zhai, H. Zhang, Simultaneous realization of anomalous reflection and transmission at two frequencies using bi-functional metasurfaces. *Sci. Rep.* **8**, 1876 (2018).
7. S. Jahani, Z. Jacob, All-dielectric metamaterials. *Nat. Nanotechnol.* **11**, 23 (2016).
8. M. Khorasaninejad, Z. Shi, A. Y. Zhu, W.-T. Chen, V. Sanjeev, A. Zaidi, F. Capasso, Achromatic metalens over 60 nm bandwidth in the visible and metalens with reverse chromatic dispersion. *Nano Lett.* **17**, 1819-1824 (2017).
9. A. Arbabi, Y. Horie, M. Bagheri, A. Faraon, Dielectric metasurfaces for complete control of phase and polarization with subwavelength spatial resolution and high transmission. *Nat. Nanotechnol.* **10**, 937 (2015).
10. S. M. Kamali, A. Arbabi, E. Arbabi, Y. Horie, A. Faraon, Decoupling optical function and geometrical form using conformal flexible dielectric metasurfaces. *Nat. Commun.* **7**, 11618 (2016).
11. Y. F. Yu, A. Y. Zhu, R. Paniagua-Domínguez, Y. H. Fu, B. Luk'yanchuk, A. I. Kuznetsov, High-transmission dielectric metasurface with 2π phase control at visible wavelengths. *Laser Photonics Rev.* **9**, 412-418 (2015).
12. L. Lewin, The electrical constants of a material loaded with spherical particles. *Journal of the Institution of Electrical Engineers-Part III: Radio and Communication Engineering* **94**, 65-68 (1947).
13. B. A. Slovick, Z. G. Yu, S. Krishnamurthy, Generalized effective-medium theory for metamaterials. *Phys. Rev. B* **89**, 155118 (2014).
14. B. Lusch, J. N. Kutz, S. L. Brunton, Deep learning for universal linear embeddings of nonlinear dynamics. *Nat. Commun.* **9**, 4950 (2018).
15. X. Lin, Y. Rivenson, N. T. Yardimci, M. Veli, Y. Luo, M. Jarrahi, A. Ozcan, All-optical machine learning using diffractive deep neural networks. *Science* **361**, 1004-1008 (2018).
16. A. D. Tranter, H. J. Slatyer, M. R. Hush, A. C. Leung, J. L. Everett, K. V. Paul, P. Vernaz-Gris, P. K. Lam, B. C. Buchler, G. T. Campbell, Multiparameter optimisation of a magneto-optical trap using deep learning. *Nat. Commun.* **9**, 4360 (2018).
17. J. Peurifoy, Y. Shen, L. Jing, Y. Yang, F. Cano-Renteria, B. G. DeLacy, J. D. Joannopoulos, M. Tegmark, M. Soljačić, Nanophotonic particle simulation and inverse design using artificial neural networks. *Science advances* **4**, eaar4206 (2018).
18. G. Cybenko, Approximation by superpositions of a sigmoidal function. *Math. Control, Signal Syst.* **2**, 303-314 (1989).
19. K. Hornik, M. Stinchcombe, H. White, Multilayer feedforward networks are universal approximators. *Neural networks* **2**, 359-366 (1989).
20. K. Hornik, M. Stinchcombe, H. White, Universal approximation of an unknown mapping and its derivatives using multilayer feedforward networks. *Neural networks* **3**, 551-560 (1990).
21. D. Liu, Y. Tan, E. Khoram, Z. Yu, Training deep neural networks for the inverse design of nanophotonic structures. *ACS Photonics* **5**, 1365-1369 (2018).

22. W. Ma, F. Cheng, Y. Liu, Deep-learning-enabled on-demand design of chiral metamaterials. *ACS nano* **12**, 6326-6334 (2018).
23. Z. Liu, D. Zhu, S. P. Rodrigues, K.-T. Lee, W. Cai, Generative model for the inverse design of metasurfaces. *Nano Lett.* **18**, 6570-6576 (2018).
24. I. Malkiel, M. Mrejen, A. Nagler, U. Arieli, L. Wolf, H. Suchowski, Plasmonic nanostructure design and characterization via Deep Learning. *Light Sci. Appl.* **7**, 60 (2018).
25. M. Decker, I. Staude, M. Falkner, J. Dominguez, D. N. Neshev, I. Brener, T. Pertsch, Y. S. Kivshar, High‐efficiency dielectric Huygens' surfaces. *Adv. Opt. Mater.* **3**, 813-820 (2015).
26. W. Liu, Y. S. Kivshar, Generalized Kerker effects in nanophotonics and meta-optics. *Opt. Express* **26**, 13085-13105 (2018).
27. P. Nikitin, A. Beloglazov, V. Kochergin, M. Valeiko, T. Ksenevich, Surface plasmon resonance interferometry for biological and chemical sensing. *Sensors and Actuators B: Chemical* **54**, 43-50 (1999).
28. R. Socher, D. Chen, C. D. Manning, A. Ng, in *Adv. Neural Inf. Process. Syst.* (2013), pp. 926-934.
29. N. Yu, P. Genevet, M. A. Kats, F. Aieta, J.-P. Tetienne, F. Capasso, Z. Gaburro, Light propagation with phase discontinuities: generalized laws of reflection and refraction. *science* **334**, 333-337 (2011).
30. H. Cai, S. Srinivasan, D. Czaplewski, A. Martinson, T. Loeffler, S. Sankaranarayanan, D. Lopez, in *High Contrast Metastructures VIII*. (International Society for Optics and Photonics, 2019), vol. 10928, pp. 109281M.
31. H. Cai, D. Czaplewski, K. Ogando, A. Martinson, D. Gosztola, L. Stan, D. López, Ultrathin transmissive metasurfaces for multi-wavelength optics in the visible. *Applied Physics Letters* **114**, 071106 (2019).
32. C. H. Chu, M. L. Tseng, J. Chen, P. C. Wu, Y. H. Chen, H. C. Wang, T. Y. Chen, W. T. Hsieh, H. J. Wu, G. Sun, Active dielectric metasurface based on phase-change medium. *Laser Photonics Rev.* **10**, 986-994 (2016).
33. Q. Wang, E. T. Rogers, B. Gholipour, C.-M. Wang, G. Yuan, J. Teng, N. I. Zheludev, Optically reconfigurable metasurfaces and photonic devices based on phase change materials. *Nature Photonics* **10**, 60 (2016).
34. Y. Zhang, J. B. Chou, J. Li, H. Li, Q. Du, A. Yadav, S. Zhou, M. Y. Shalaginov, Z. Fang, H. Zhong, C. Roberts, P. Robinson, B. Bohlin, C. Rios, H. Lin, M. Kang, T. Gu, J. Warner, V. Liberman, K. Richardson, J. Hu, Extreme Broadband Transparent Optical Phase Change Materials for High-Performance Nonvolatile Photonics. *arXiv preprint arXiv:1811.00526*, (2018).
35. J. Tian, Q. Li, J. Lu, M. Qiu, Reconfigurable all-dielectric antenna-based metasurface driven by multipolar resonances. *Opt. Express* **26**, 23918-23925 (2018).

**Acknowledgements**

The work is funded under Defense Advanced Research Projects Agency Defense Sciences Office (DSO) Program: EXTREME Optics and Imaging (EXTREME) under Agreement No. HR00111720029.

**Author contributions**

S.A. and H.Z. conceived the methodology concepts. S.A., L.Z. and B.Z. contributed to setting up the machine learning framework and designing the neural networks. H.T., H.L. and J.D. assisted in data collection and processing. J.H. and H.Z. supervised and coordinated the project. S.A., C.F., J.H. and H.Z. wrote the manuscript. All authors contributed to technical discussions regarding this work.